\documentclass[prl,twocolumn,twoside,showpacs,floatfix]{revtex4}
\usepackage{graphics,amssymb,amsmath,epsf}
\epsfclipon

\begin{document}
\title{Anomalous Binder Cumulant and Lack of Self-averageness \\
in Systems with Quenched Disorder}

\author{Hyunsuk Hong}
\affiliation{Department of Physics and Research Institute of Physics and
Chemistry, Chonbuk National University, Jeonju 561-756, Korea}

\author{Hyunggyu Park}
\affiliation{School of Physics, Korea Institute for Advanced Study, Seoul
130-722, Korea}

\author{Lei-Han Tang}
\affiliation{Department of Physics, Hong Kong Baptist University, Kowloon Tong,
Hong Kong SAR, China}

\date{\today}

\begin{abstract}
The Binder cumulant (BC) has been widely used for locating the phase transition
point accurately in systems with thermal noise. In systems with quenched
disorder, the BC may show subtle finite-size effects due to large
sample-to-sample fluctuations. We study the globally coupled Kuramoto model of
interacting limit-cycle oscillators with random natural frequencies and find an
anomalous dip in the BC near the transition. We show that the  dip is related
to non-self-averageness of the order parameter at the transition. Alternative
definitions of the BC, which do not show any anomalous behavior regardless of the 
existence of non-self-averageness, are proposed. 

\end{abstract}
%05.45.Xt: Synchronization, coupled oscillators
%89.75.-k: Complex systems
%05.45.-a Nonlinear dynamics and nonlinear dynamical systems
\maketitle
\vspace{5cm}

%\section {I. Introduction}

The characterization of phase transitions relies mainly on the singularity 
structure of physical quantities at the transition, which can be quantified by critical
exponent values. In numerical efforts, the accuracy of the estimated exponents
heavily depends on the precision of locating the phase transition point.
In the case of most thermal systems, 
the Binder cumulant (BC) is widely believed to provide one of the most 
accurate tools for estimating the transition
point~\cite{ref:BC1,ref:BC11,ref:BC2}. The critical BC value at the transition
is also believed to be universal, even though there is still controversy over 
its universality~\cite{ref:BC3}.

In some complex systems~\cite{ref:BC4}, the BC shows an anomalous negative dip
in finite systems, which represents a rugged landscape (multi-peak structure)
in the probability distribution function (PDF) of the order parameter. 
Great care is required in analyzing numerical data to see whether the dip will
vanish in the thermodynamic limit. If it does, the negative dips in the finite
systems can be attributed to long-living metastable states. Otherwise, a
nonvanishing negative dip usually implies that the transition is not
continuous, but is of the first order.

In systems with quenched disorder, the disorder fluctuation may also generate
an anomalous negative dip in the conventional BC, which is defined as the ratio
of the disorder-averaged moments of the order parameter. In this case, the
negative dip may be related to the non-self-averageness (NSA) of the order
parameter, which usually implies an extended and/or multi-peak structure in the
disorder-averaged PDF~\cite{ref:BC5}.

We consider a typical nonequilibrium dynamical system with quenched disorder,
such as the Kuramoto model of interacting limit-cycle oscillators with random
natural frequencies~\cite{ref:Kuramoto}. The dynamic synchronization transition
is dominated by space-time fluctuations of the order parameter. The quenched
disorder is, by definition, perfectly correlated in the time direction, so it
may generate strong disorder fluctuations similar to quantum systems with
random defects~\cite{ref:QS}.  In fact, we recently showed that the disorder 
fluctuation was anomalously strong near the synchronization
transition~\cite{ref:Hong,ref:HCPT}.

We take the globally coupled Kuramoto model, which can be solved analytically to
some extent. The model is defined by the set of equations of motion
\begin{equation}
\frac{d\varphi_i}{dt} = \omega_i - \frac{K}{N}\sum_{j=1}^{N}\sin(\varphi_i -
\varphi_j), \label{eq:model}
\end{equation}
where $\varphi_i$ represents the phase of the $i$th limit-cycle oscillator
$(i=1,2,\cdots,N)$.  The first term $\omega_i$ on the right-hand side denotes the
natural frequency of the $i$th oscillator, where $\omega_i$
is assumed to be randomly distributed according to the Gaussian distribution function $g(\omega)$
characterized by the correlation $\langle\omega_i \omega_j\rangle =2\sigma\delta_{ij}$ and zero
mean $(\langle \omega_i \rangle=0)$.

We note that the natural frequency $\omega_i$ plays the role of {\it ``quenched
disorder"}. The second term of Eq.~(\ref{eq:model}) represents global
(all-to-all) coupling with equal coupling strength $K/N$. The sine coupling
form is the most general representation of the coupling in the lowest order of
the complex Ginzburg-Landau (CGL) description\cite{ref:Kuramoto}, and its
periodic nature is generic in limit-cycle oscillator systems. We consider the
ferromagnetic coupling ($K>0$), so the neighboring oscillators favor their
phase difference being minimized. The scattered natural frequencies and the coupling
of the oscillators compete with each other. When the coupling becomes strong enough
to overcome the dispersion of natural frequencies, macroscopic regions
in which the oscillators are synchronized by sharing a coupling-modified
common frequency $\Omega=0$ may emerge.

Collective phase synchronization is conveniently described by the complex order
parameter defined by
\begin{equation}
\Delta e^{i\theta} \equiv
\frac{1}{N}\sum_{j=1}^{N} e^{i\varphi_j},
\label{eq:def_phaseorder}
\end{equation}
where the amplitude $\Delta$ measures the phase synchronization and $\theta$
indicates the average phase. When the coupling is weak, each oscillator tends
to evolve with its own natural frequency, resulting in the fully random
desynchronized phase ($\Delta=0$). As the coupling increases, some oscillators
with $\omega_i\approx 0$ become synchronized, and their phases $\phi_i$ start to
show some ordering ($\Delta> 0$).

Equation~(\ref{eq:model}) can be simplified to $N$ decoupled equations
\begin{equation}
\frac{d\varphi_i}{dt}= \omega_i - K\Delta\sin(\varphi_i - \theta),
\label{eq:decoupled}
\end{equation}
where $\Delta$ and $\theta$ are to be determined by imposing self-consistency. 
In the steady state ($t\rightarrow\infty$), the
self-consistency equation reads
\begin{equation}
\Delta=a(K\Delta)-b(K\Delta)^3 + {\cal{O}}(K\Delta)^5
\end{equation}
with $a=(\pi/2) g(0)$ and $c=-(\pi/16)g''(0)$~\cite{ref:Kuramoto}.
%which yields $K_c=2/\pi g(0)$~\cite{ref:Kuramoto}.
This equation has a nontrivial solution only when $K>K_c=1/a$:
\begin{equation}
\Delta \sim (K-K_c)^{\beta}
\end{equation}
with $\beta=1/2$. We note that the exponent $\beta=1/2$ corresponds to the mean
field (MF) value for systems of locally coupled
oscillators~\cite{ref:Kuramoto}.

Now, we perform numerical integrations of Eq.~(\ref{eq:model}) by using Heun's
method~\cite{ref:Heun} for various system sizes of $N=200$ to $12800$. For a given
distribution of disorder $\{\omega_i\}$, we average over time in the steady
state after some transient time.  After the time average, we also average over
disorder. Typically, we take the time step $\delta t=0.05$, the maximum number
of time steps $N_t=4\times 10^4$, and the number of samples $N_s=100 \sim 1000$. For
convenience, we set $2\sigma=1$ (unit variance); then, the corresponding
critical parameter value is $K_c=\sqrt{8/\pi}=1.595769...$.

\begin{figure}
%\centering{\resizebox*{!}{5.0cm}{\includegraphics{gl_Delta.eps}}}
%\centering{\resizebox*{!}{5.0cm}{\includegraphics{gl_Delta_0905.eps}}}
\centering{\resizebox*{!}{5.0cm}{\includegraphics{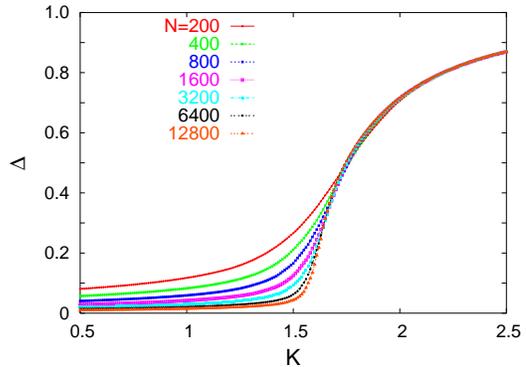}}}
\caption{Phase synchronization order parameter $\Delta$ versus the coupling
strength $K$ for various system sizes $N$.} \label{fig:gl_M}
\end{figure}

Figure~\ref{fig:gl_M} shows the behavior of the phase synchronization order
parameter $\Delta$ against the coupling strength $K$ for various system sizes
$N$.  In the weak coupling region ($K\lesssim 1.6$), we find that the order
parameter approaches zero as $\Delta\sim N^{-1/2}$, which is a characteristic
of the fully random phase. In the strong coupling region $(K\gtrsim 1.6)$,
$\Delta$ saturates to a finite value, indicating a phase transition at
$K\approx 1.6$ in the thermodynamic limit $(N\rightarrow \infty)$, which is
consistent with the analytic result.
%%%%%%%%%%%%%%%%%%%%%%%%%%%%%%%%%%%%%%%%%%%%%%%%%%%%%%%%%%%%%%%%%%%%%%%%%%%%%%%%%%%%
%%%%%%%%%%%%%%%%%% Fig. 1 (Phase order parameter Delta  %%%%%%%%%%%%%%%%%%%%%%%%%%%%
%%%%%%%%%%%%%%%%%%%%%%%%%%%%%%%%%%%%%%%%%%%%%%%%%%%%%%%%%%%%%%%%%%%%%%%%%%%%%%%%%%%%
%%%%%%%%%%%%%%%%%%%%%%%%%%%%%%%%%%%%%%%%%%%%%%%%%%%%%%%%%%%%%%%%%%%%%%%%%%%%%%%%%%%%

To pin down the transition point $K_c$ precisely, we use the Binder cumulant
method~\cite{ref:BC1,ref:BC11}. The fourth-order cumulant of the order
parameter, the Binder cumulant (BC), is defined in thermal systems as
\begin{equation}
%B_{\Delta} =  1-\frac{[\langle \Delta^4 \rangle ]}{3[\langle \Delta^2  \rangle]^2}
B_{\Delta} =  1-\frac{\langle \Delta^4 \rangle }{3\langle \Delta^2  \rangle^2},
\label{eq:gl_BM}
\end{equation}
where $\langle \cdots\rangle$ represents the thermal (time) average. In systems
with quenched disorder, on the other hand, we should consider the disorder
average besides the thermal one. We may first consider the BC as the
disorder-averaged moment ratio~\cite{ref:BC2,ref:BC5,ref:SG}
\begin{equation}
B_{\Delta}^{(1)} =  1-\frac{[\langle \Delta^4 \rangle ]}{3[\langle \Delta^2
\rangle]^2}, \label{eq:gl_BM_old}
\end{equation}
where $[\cdots]$ denotes the disorder average, i.e., the average over
different realizations of $\{\omega_i\}$.

\begin{figure}
%\centering{\resizebox*{!}{5.0cm}{\includegraphics{gl_Binder_old.eps}}}
%\centering{\resizebox*{!}{5.0cm}{\includegraphics{gl_Binder_old_0905.eps}}}
\centering{\resizebox*{!}{5.0cm}{\includegraphics{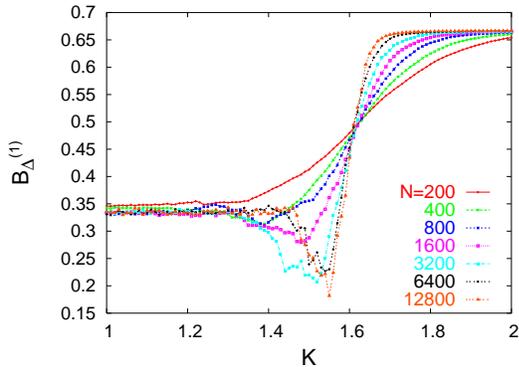}}}
\caption{Binder cumulant $B^{(1)}_{\Delta}$, defined by Eq.~(\ref{eq:gl_BM_old}), of the 
phase synchronization order parameter $\Delta$.}
\label{fig:gl_BM_old}
\end{figure}

Figure~\ref{fig:gl_BM_old} displays $B_{\Delta}^{(1)}$ as a function of the
coupling strength $K$ for various system sizes $N$. In the region of weak
coupling $(K\rightarrow 0)$, we expect the random nature of the oscillator
phases $\{\phi_i\}$ to yield an asymmetric Poisson-like probability
distribution function (PDF) characterized by $P(\Delta)\sim \Delta
\exp(-c\Delta^2)$ with a constant $c$, which leads to $B_{\Delta}^{(1)}=1/3$.
On the other hand, in the strong-coupling region, the PDF becomes a
$\delta$-like function with a very narrow variance, which leads to
$B_{\Delta}^{(1)}=2/3$. The numerical data in Fig.~2 are consistent with our
predictions.

However, near the transition, the $B_{\Delta}^{(1)}$ shows a big anomalous
{\it ``dip"} on the desynchronized side. As the system size increases, the dip
develops initially with a broad width and then becomes sharper and also deeper.
The dip's position moves toward the transition point.  The crossing points 
seem to nicely converge to the critical point $K_c=\sqrt{8/\pi}$.
However, as the system size increases, the presence of the dip starts to hinder
us in locating the critical point accurately.

\begin{figure}
%\centering{\resizebox*{!}{5.0cm}{\includegraphics{gl_A_1.eps}}}
%\centering{\resizebox*{!}{5.0cm}{\includegraphics{gl_A_0905.eps}}}
\centering{\resizebox*{!}{5.0cm}{\includegraphics{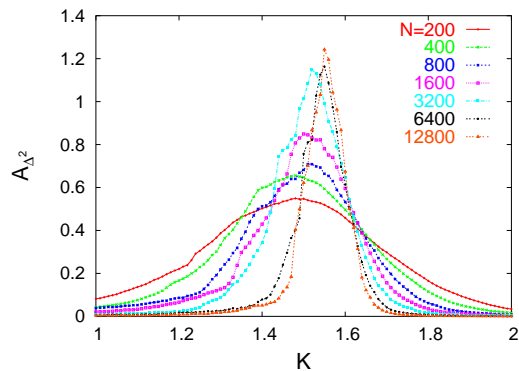}}}
\caption{Disorder fluctuation $A_{\Delta^2}$ defined by Eq.~(\ref{eq:A}).}
\label{fig:gl_A}
\end{figure}

In this Letter, we explain why the dip develops in this system and propose
alternative definitions of the Binder cumulant that do not show any dip in the
same system. We measure the disorder (sample-to-sample) fluctuations defined as
\begin{equation}
A_{\cal O}=\frac{[\langle {\cal O}\rangle^2]}{[\langle{\cal O}\rangle]^2}-1,
\label{eq:A}
\end{equation}
where ${\cal O}$ is any observable, such as $\Delta$ and $\Delta^2$, in a system. 
This quantity is positive definite and is supposed to vanish in the
thermodynamic limit in {\em self-averaging} systems and to remain finite in
non-self-averaging systems~\cite{ref:NSA,ref:SG}. As one can see in
Fig.~\ref{fig:gl_A}, the disorder fluctuation $A_{\Delta^2}$ is quite sizable
in the range of $K$ where the dip appears ($A_\Delta$ shows a similar
behavior). In other words, the $B_{\Delta}^{(1)}$ shows a dip where the system
is not well self-averaged.  A careful finite-size analysis on $A_{\Delta^2}$
reveals that it vanishes as $\sim N^{-1}$ away from criticality, but saturates
to a finite value at criticality. The non-self-averageness at criticality is
not surprising because the quenched randomness in natural frequencies should be
relevant at this transition.

%%%%%%%%%%%%%%%%%%%%%%%%%%%%%%%%%%%%%%%%%%%%%%%%%%%%%%%%%%%%%%%%%%%%%%%%%%%%%%%%%%%%
%%%%%%%%%%%%%%%%%%%%% Fig. 2 Binder cumulant of  Delta  %%%%%%%%%%%%%%%%%%%%%%%%%%%%
%%%%%%%%%%%%%%%%%%%%%%%%%%%%%%%%%%%%%%%%%%%%%%%%%%%%%%%%%%%%%%%%%%%%%%%%%%%%%%%%%%%%
%%%%%%%%%%%%%%%%%%%%%%%%%%%%%%%%%%%%%%%%%%%%%%%%%%%%%%%%%%%%%%%%%%%%%%%%%%%%%%%%%%%%
%To see the dip feature shown in Fig.~\ref{fig:gl_BM_old} is real fact, we now try to measure the

%%%%%%%%%%%%%%%%%%%%%%%%%%%%%%%%%%%%%%%%%%%%%%%%%%%%%%%%%%%%%%%%%%%%%%%%%%%%%%%%%%%%
%%%%%%%%%%%%%%%%%% Fig. 3   PDF  %%%%%%%%%%%%%%%%%%%%%%%%%%
%%%%%%%%%%%%%%%%%%%%%%%%%%%%%%%%%%%%%%%%%%%%%%%%%%%%%%%%%%%%%%%%%%%%%%%%%%%%%%%%%%%%
\begin{figure}
%\centering{\resizebox*{!}{5.0cm}{\includegraphics{gl_PDF_M.eps}}}
%\centering{\resizebox*{!}{5.0cm}{\includegraphics{gl_PDF_M_0907.eps}}}
\centering{\resizebox*{!}{5.0cm}{\includegraphics{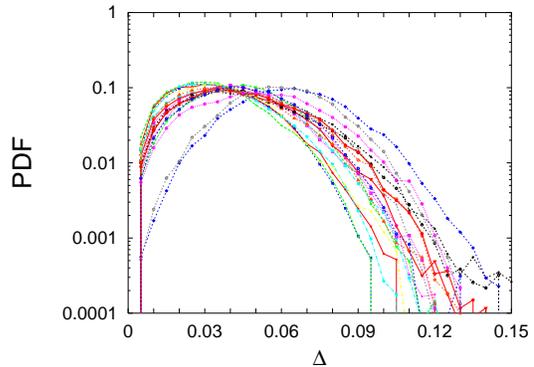}}}
\caption{Probability distribution function (PDF) at $K=1.5$ and $N=12800$ in
the steady state.  Each curve corresponds to one of 20 independent samples.}
\label{fig:PDF}
\end{figure}
%%%%%%%%%%%%%%%%%%%%%%%%%%%%%%%%%%%%%%%%%%%%%%%%%%%%%%%%%%%%%%%%%%%%%%%%%%%%%%%%%%%%
%%%%%%%%%%%%%%%%%%%%%%%%%%%%%%%%%%%%%%%%%%%%%%%%%%%%%%%%%%%%%%%%%%%%%%%%%%%%%%%%%%%%
%%%%%%%%%%%%%%%%%% Fig. 4   New Binder cumulant of Delta  %%%%%%%%%%%%%%%%%%%%%%%%%%
%%%%%%%%%%%%%%%%%%%%%%%%%%%%%%%%%%%%%%%%%%%%%%%%%%%%%%%%%%%%%%%%%%%%%%%%%%%%%%%%%%%%
\begin{figure}
%\centering{\resizebox*{!}{5.0cm}{\includegraphics{gl_Binder_new.eps}}}
%\centering{\resizebox*{!}{5.0cm}{\includegraphics{gl_Binder_new_0905.eps}}}
\centering{\resizebox*{!}{5.0cm}{\includegraphics{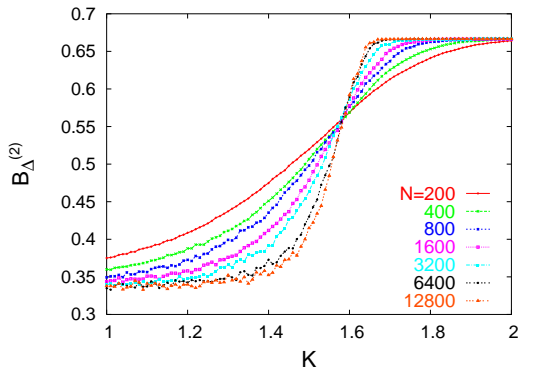}}}
\caption{Binder cumulant $B^{(2)}_{\Delta}$ defined  by
Eq.~(\ref{eq:gl_BM_new}). Note that the dip shown in $B^{(1)}_{\Delta}$
disappears. }
%yields less size effects comparing with the measurement given by
%Eq.~(\ref{eq:gl_BM_old}).}
\label{fig:gl_BM_new}
\end{figure}
%%%%%%%%%%%%%%%%%%%%%%%%%%%%%%%%%%%%%%%%%%%%%%%%%%%%%%%%%%%%%%%%%%%%%%%%%%%%%%%%%%%%

Strong disorder fluctuations may cause non-negligible spreading of the {\em
effective}  coupling constants over different realizations of
disorder~\cite{ref:NSA}. 
%In order to measure the proper characteristic of the
%PDF (e.g.~Binder cumulant), one should shift and coincide the effective
%coupling constants for all samples before averaging the PDF over disorder
%realizations. 
%% Notes by Leihan: At this point it is hard to say what is a ``proper''
%% way to introduce the Binder cumulant. Essentially, the issue is how
%% broad the PDF is, after averaging over the disorder. Since higher moments
%% are more sensitive to the tails of the PDF, and the tail may scale differently
%% from the middle (which may indeed be the case here), the BC may yield different
%% results depending on the averaging procedure.
Figure~\ref{fig:PDF} shows for 20 independent samples, the PDF of $\Delta$ 
just below the transition and obtained from the time series of $\Delta$ after
the system had reached the steady state. Indeed, a large part of the
sample-to-sample variations can be interpreted as a shift in the $K_c$ of
individual samples. The two quantities ${[\langle\Delta^2\rangle]}$ and
$[\langle\Delta^4\rangle]$ in Eq. (\ref{eq:gl_BM_old}) can be considered 
as the second and the fourth moments of the disorder-averaged PDF, which is
much broader than the individual PDFs near the transition.  One can easily
see that broadening yields a larger value for the ratio
$[\langle\Delta^4\rangle]/[\langle\Delta^2\rangle]^2$ and, hence, a smaller BC.
The effect is particularly pronounced on the small $K$ side of the transition,
where $\Delta$ itself is small, in which case a shift in $K_c$ has a stronger 
influence on the moments.

An alternative definition for the Binder cumulant for systems with quenched 
disorder (especially non-diminishing disorder fluctuations) is\cite{ref:QS,ref:HH}
\begin{equation}
B^{(2)}_{\Delta} =  1-\Biggl [\frac{\langle \Delta^4 \rangle}{3\langle \Delta^2
\rangle^2} \Biggr ]. \label{eq:gl_BM_new}
\end{equation}
We note that the disorder average is performed over the ratio of the time-averaged 
moments. The moment ratio is calculated for each sample first and, is then
averaged over disorder.  It is clear that this definition of the Binder cumulant
should eliminate the most dominant contribution from the disorder fluctuations,
i.e.,~the anomaly caused by the spreading of the effective coupling constants.
This definition has been adopted mostly in quantum disorder systems, where 
strong disorder fluctuations are anticipated~\cite{ref:QS}.
Figure~\ref{fig:gl_BM_new} displays $B^{(2)}_{\Delta}$ versus $K$. We note that
the dip shown in Fig.~\ref{fig:gl_BM_old} disappears and that the crossing points
nicely converge to $K_c$, implying that $B^{(2)}_{\Delta}$ should serve better
for locating the transition point than the conventional one, which is confirmed
numerically (not shown here).

%%%%%%%%%%%%%%%%%%%%%%%%%%%%%%%%%%%%%%%%%%%%%%%%%%%%%%%%%%%%%%%%%%%%%%%%%%%%%%%%%%%%
%%%%%%%%%%%%%%%%%% Fig. 5   Newer Binder cumulant of Delta  %%%%%%%%%%%%%%%%%%%%%%%%%%
%%%%%%%%%%%%%%%%%%%%%%%%%%%%%%%%%%%%%%%%%%%%%%%%%%%%%%%%%%%%%%%%%%%%%%%%%%%%%%%%%%%%
\begin{figure}
%\centering{\resizebox*{!}{5.0cm}{\includegraphics{gl_Binder_newer.eps}}}
%\centering{\resizebox*{!}{5.0cm}{\includegraphics{gl_Binder_Park_0905.eps}}}
\centering{\resizebox*{!}{5.0cm}{\includegraphics{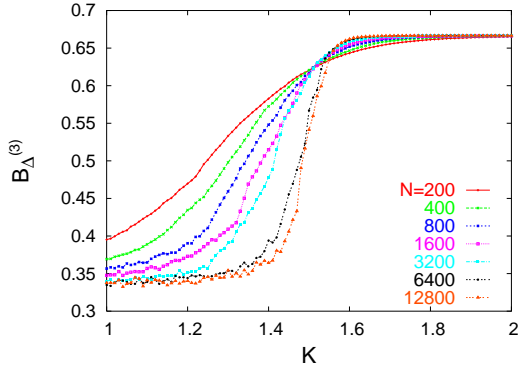}}}
\caption{Binder cumulant $B^{(3)}_{\Delta}$ defined  by
Eq.~(\ref{eq:gl_BM_newer}).  }
%yields less size effects comparing with the measurement given by
%Eq.~(\ref{eq:gl_BM_old}).}
\label{fig:gl_BM_newer}
\end{figure}
%%%%%%%%%%%%%%%%%%%%%%%%%%%%%%%%%%%%%%%%%%%%%%%%%%%%%%%%%%%%%%%%%%%%%%%%%%%%%%%%%%%%

Yet another definition of the Binder cumulant is
\begin{equation}
B_{\Delta}^{(3)} =  1-\frac{[\langle \Delta^4 \rangle ]}{3[\langle \Delta^2
\rangle^2]}. \label{eq:gl_BM_newer}
\end{equation}
We expect that $B_{\Delta}^{(3)}$ may also behave smoothly near the transition
because it does not involve disorder fluctuation terms such as $[\cdots]^2$
included in $B_{\Delta}^{(1)}$. Figure~\ref{fig:gl_BM_newer} displays
$B^{(3)}_{\Delta}$ versus $K$. As expected, we find no anomalous behavior in
$B_{\Delta}^{(3)}$. We can directly relate $B_{\Delta}^{(1)}$ and
$B_{\Delta}^{(3)}$ through the disorder fluctuation $A_{\Delta^2}$.  Simple
algebra leads to
\begin{equation}
B^{(1)}_\Delta=B^{(3)}_\Delta -(1-B^{(3)}_\Delta)A_{\Delta^2}.
\end{equation}
As the disorder fluctuation $A_{\Delta^2}$ becomes larger, $B^{(1)}_\Delta$
shows a bigger dip. This explains quantitatively the size and the location of the
dip in $B^{(1)}_\Delta$. The critical value of $B^{(3)}_\Delta$ ($\approx 2/3$)
provides additional information on the temporal variations of $\Delta$. 
One can show that
$3B^{(3)}_\Delta= 2 - [\langle\delta\Delta^2\rangle]/[\langle\Delta^2\rangle^2]$, 
where $\langle\delta\Delta^2\rangle=\langle\Delta^4\rangle-\langle\Delta^2\rangle^2$.
%% How is \delta\Delta defined?
Our numerical result indicates that the relative temporal fluctuations are
almost negligible even at criticality. In this case, $B^{(3)}_\Delta$ is
not practically useful in locating the transition point accurately.

In summary, we studied Binder cumulants in the quenched disorder system.
For the  Kuramoto model, we found that the conventionally defined BC shows a
big  anomalous dip near the transition. This dip is shown to be directly
related to the disorder fluctuation (non-self-averageness). Alternative
definitions of the BC, which did not show any anomalous behavior were proposed
and may be useful in locating the transition point accurately in general systems
with quenched disorder.

\section*{Acknowledgments}
This work was supported by research funds of Chonbuk National University (2004) and the 
Korea Research Foundation Grant (MOEHRD) (R14-2002-059-01000-0) (HH), and by the 
Research Grants Council of the HKSAR under project 2017/03P and Hong Kong Baptist 
University under project FRG/01-02/II-65 (LHT).

\end{document}